# Design Study on Medium beta SC Half-Wave Resonator at IMP*

WU An-Dong(吴安东)[1,2;1)]   ZHANG Sheng-Hu(张生虎)[1]   YUE Wei-Ming(岳伟明)[1]   LI Yong-Ming(李永明)[1]
JIANG Tian-Cai(蒋天才)[1]   WANG Feng-Feng(王锋锋)[1]   ZHANG Sheng-Xue(张升学)[1]   HUANG Ran(黄燃)[1,2]
HE Yuan(何源)[1]   ZHAO Hong-Wei(赵红卫)[1]

[1] Institute of Modern Physics, Chinese Academy of Sciences, Lanzhou 730000, China

[2] University of Chinese Academy of Sciences, Beijing 100049, China

**Abstract:** A superconducting half-wave resonator has been designed with the frequency of 325 MHz and beta of 0.51. Different geometry parameters and shapes of inner conductors (racetrack, ring-shape and elliptical-shape) were optimized to decrease the peak electromagnetic fields to obtain higher accelerating gradients and minimize the dissipated power on the cavity walls. To suppress the operation frequency shift caused by the helium pressure fluctuations and maximize the tuning ranges, the frequency shifts and mechanical properties were studied on the electric and magnetic areas separately. At the end, the helium vessel was also designed to keep the mechanical structure as robust as possible. The fabrication and test of the prototype will be completed in the beginning of 2016.

## 1 Introduction

The production development of the 162.5 MHz half-wave resonators (HWR) with the optimal beta of 0.101 has achieved a great success and revealed excellent performance on the High Current Proton Superconducting Linac for C-ADS Injector II at Institute of Modern Physics [1-2]. Based on our experience and experiment of superconducting cavity R&D, a medium beta prototypal half-wave resonator was proposed. Because of its wide longitude acceptance and bigger beam pipes, it can be utilized to propel the proton or heavy ion beams from the beta (relative velocity to the light) of 0.35 to 0.65, which is the medium accelerating section that connects low energy sections with the elliptical cavity zones.

In the recent years, a great progress of the HWRs, with the medium beta of 0.53, has been achieved in the construction of FRIB at MSU [3]. Other types of transmission superconducting resonators, for example, the single spoke with the beta of 0.515 for Project X and the double spoke with the beta of 0.5 for ESS, also have been studied for medium energy section [4-5]. Compared with the structure of the spoke, the HWR have simpler geometry, compact volume and more outstanding mechanical stability for operation. Thus, it attracts us to research its application on the medium beta section for the high current and continuous wave mode linear accelerators, such as the more compact ADS facility [6].

In our research, an elliptical inner conductor was proposed to optimize the distribution of surface field, $E_{peak}/E_{acc}$ of 4.03 and $B_{peak}/E_{acc}$ of 7.08 (mT/MV/m) were achieved, where $E_{acc}$ is the accelerating gradient defined by the voltage gain over the efficient length of $\beta\lambda$. In addition, the frequency sensitivity with the helium vessel was optimized at 5.6 Hz/mbar.

## 2 RF design

In the low beta section, the HWR010s operate at relatively low accelerating gradient of 5MeV/m with the voltage gains of 0.78 MeV, limited by the beam dynamics of longitude mismatching [7]. However, the higher accelerating gradient is usually quested for medium beta section, in order to decrease the construction cost by reducing the length of linac and numbers of affiliated machines (power couplers, tuners and RF generators). Higher accelerating voltage also accompanies with higher surface peak fields and dozens of dissipated watts. Moreover, higher peak surface fields possibly lead to field emission and superconducting quench. Especially, the thermal broken constantly occurred in our vertical test, when the peak magnetic field exceeds 90 mT [8]. Thus, the peak fields are limited with the $E_{peak}$ of 35

---

*Supported by the important Directional Program of the Chinese Academy of Sciences (Y115210YQO), 973 Project (Y437030KJO).

1) E-mail: antonwoo@impcas.ac.cn

MV/m and $B_{peak}$ of 60 mT.

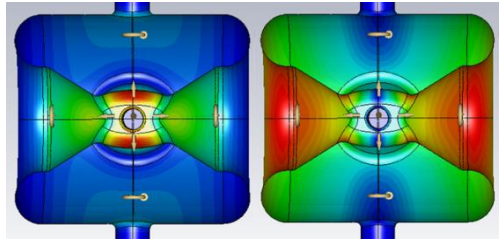

Fig. 1. The surface electric (left) and magnetic (right) field distribution of the elliptical center inner conductor HWR. The field amplitude decreases as the color changes from red to blue.

The main task of RF design aims at reducing the normalized magnetic field and dissipated power. The RF property simulations were complete by the CST Microwave Studio [9]. The geometry of HWR, with TEM-alike class structure, consists of inner and outer conductors. The electric field concentrates on the center of inner conductor near the beam pipes, and magnetic field encircles around the inner conductor of two short domes, as illustrated in Fig. 1. The RF properties mainly depend on the shape of inner conductor and parameters of the ends dome. In order to identify the proper shape for obtaining a uniformed distributed fields and higher shunt impedances, three types of inner conductors were taken into optimizations, which named after the its cross profiles with ring shaped (RS), race track (RT) and elliptical shaped (ES), as shown in Fig. 2. Additionally, the cavity aperture radius is 25 mm for the beam dynamics consideration.

**2.1 Choosing of structure**

The quadrupole asymmetry effects of the inner center conductor are comprehensively analyzed for the low beta coaxial resonators, especially the comparison between the race track and ring shaped inner center conductors [10-11]. Besides the advantages of better symmetry of accelerating field, which is the results of symmetrical geometry along with the beam axial line, the center conductor with ring shaped can also provide higher $R/Q$ and lower $B_{peak}/E_{acc}$. In the cavity length axial direction, because of the loft transition from a circle section in the center conductor to the round base of inner dome, the ring structure can produce a magnetic field distribution of more flatness. However, the surface of ring shaped has a more sharped variation than that of the race track, so it is hard for ring shaped structure to get lower normalized surface electric field compared with race track. In order to obtain lower normalized electromagnetic fields and also higher $R/Q$, an elliptical shaped center conductor was proposed, which can combine the advantages of the race track with that of ring shaped structures.

In the process of the RF design, the total length of cavity was varied to keep the fundament mode frequency fixed at 325 MHz, and the inner length of iris-to-iris was adjusted to keep the optimized beta fixed at 0.51. Generally, every RF parameter has dependence on the each dimension of geometry with different sensitivity. As shown in Fig. 1, the electric and magnetic field distributions clearly are separated into two areas. Thus, the structure parameters can be divided into magnetic groups $B_{peak}/E_{acc}$ and electric groups $E_{peak}/E_{acc}$ to optimize in the relatively independent progress.

The ring shaped can be viewed as a special case of the elliptical shaped that its semi major axis (along with the cavity length) is equal to the semi minor axis (vertical to the beam line). Therefore, we just illustrate the design processes of the race track and elliptical shaped geometry in the next section.

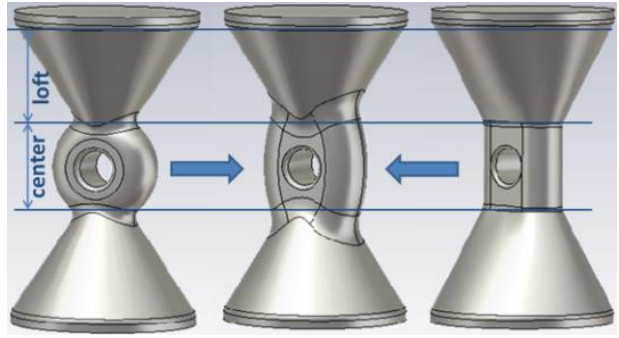

Fig. 2. Three types of inner conductors: ring shaped (left), elliptical shaped (medium) and race track (right).

**2.2 Magnetic field domains**

On the magnetic dominant area, the main structure parameters of the cavity top radius (CVTR) and inner conductor top radius (ICTR) have significant effects on the $B_{peak}/E_{acc}$ and $R/Q*G$, as illustrated in Fig .3. Four of these plots clearly demonstrate that the elliptical shaped center conductor has the better value of $B_{peak}/E_{acc}$ and $R/Q*G$ than the race track, when the CVTR and ICTR at the same dimensions.

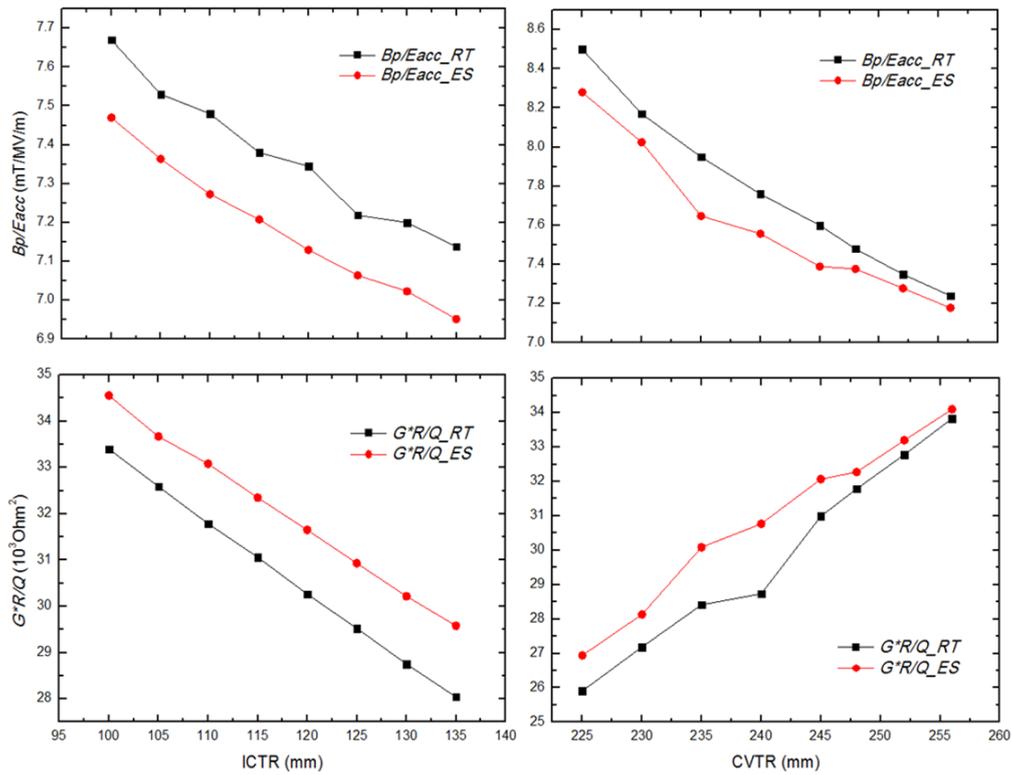

Fig. 3. Dependence of normalized magnetic fields (the upper two) and $R/Q*G$ (the lower two) on the values of ICTR and CVTR. Two types of center inner conductors are used: the red for the elliptical shaped (ES) and the black for the race track (RT).

For the two type of inner conductors, their top radiuses were fixed at 110 mm, when we swept the value of CVTR from 225 to 255 mm. When the cavity top radii increase, the tendency of $B_{peak}/E_{acc}$ and $G*R/Q$ are improved, with $B_{peak}/E_{acc}$ decreased about 12% and $G*R/Q$ increased about 30%. The increasing of cavity top radiuses provides a positive effect on RF properties, especially for 4 K operation , large variation in $G*R/Q$ will have a dramatic difference in the dissipated power. While, the larger cavity top radius will increase the cost of fabrication and cooling equipment. CVTR is chosen to be at 248 mm as a compromise.

The sweeping of ICTR illustrates that $B_{peak}/E_{acc}$ and $G*R/Q$ is decaying down about 19% and 17%. The surface current density can be reduced by increasing the inner surface areas, which can be utilized to minimize the surface magnetic fields, and decreasing the volume of cavities will have a negative compact on $G*R/G$, when the inner conductor radiuses are increasing. At last, 124 mm is chose for the top radius of inner conductor for two types of center conductor at such a contradictory RF property.

**2.3 Inner center conductors**

For the structure parameters of inner center conductor, the variation of the dimensions that along the cavity length, i.e., the height of the race track and the semi major axis of the elliptical shaped, do not have significant influence on RF characteristics. Therefore, the optimization of electric areas mainly focused on the width and thickness of the center inner conductors, as depicted in the Fig. 4.

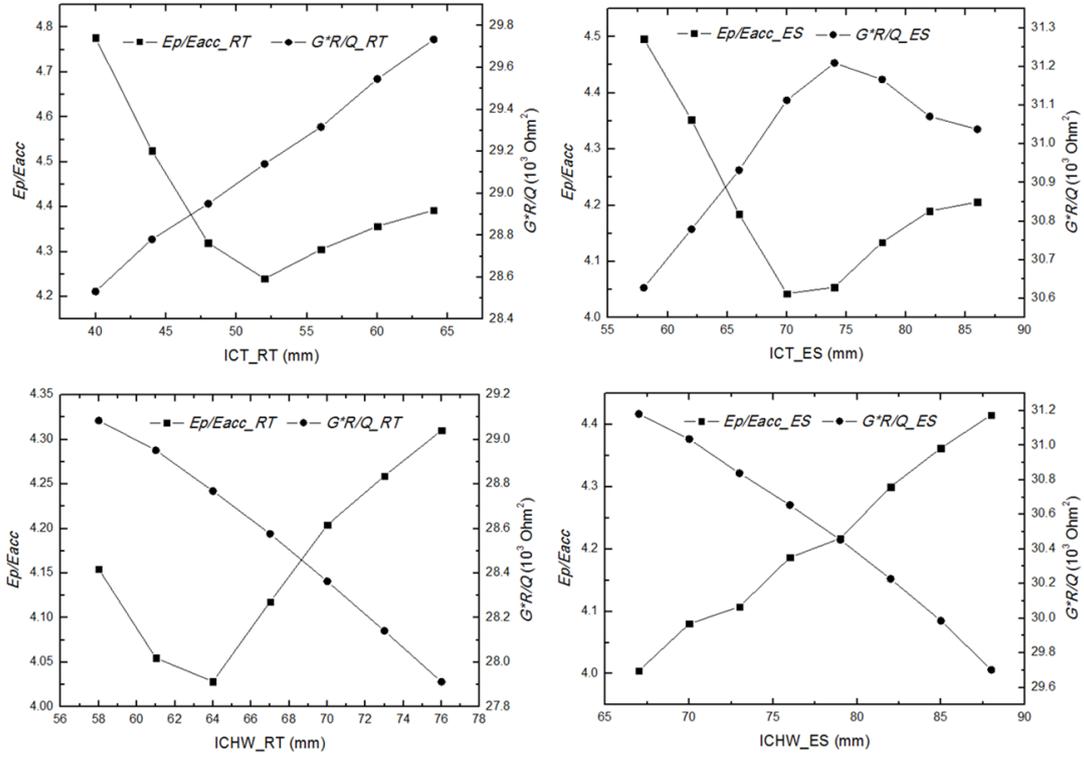

Fig. 4. Dependence of $E_{peak}/E_{acc}$ and $R/Q*G$ on parameter of race track and ring shaped. The top two are for the beam thickness of inner conductors. The bottoms two are for the half width of race track (left) and the semi minor axis of the elliptical shaped (right).

The variation of ICT have similar compacts on the $E_{peak}/E_{acc}$ for two types of center conductors, the $E_p / E_{acc}$ firstly decrease and then increase with the increasing ICT. Increasing ICT can obtain higher $G*R/Q$ for the race track conductor, but it has opposite tendency for elliptical shaped conductor when the value of ICT exceeds 75 mm.

Increasing ICHW can reduce the volumes that stored electromagnetic energy, have the negative dependencies on $G*R/Q$ values, but the differences are tiny around 6% for two types of conductors. When sweeping the width of race track, there will exist the minimum value of the $E_{peak}/E_{acc}$ on the middle point.

**2.4 Results and comparisons**

Combined advantage of race track and ring shaped, the elliptical shaped can obtain the lower normalized surface fields and higher *R/Q*, as shown in Table 1. With the same magnetic areas structures, the $E_{peak}/E_{acc}$ of the ring shaped (RS) is about 18% higher than the rest. The *R/Q* of the race track is about 10% lower than others, and also lower $E_{peak}/E_{acc}$. Therefore, we chose the elliptical shaped center conductor as the final RF structure for further study and fabrication.

Table 1. Simulation results of three types of center inner conductors for HWR051 at IMP and the parameters of HWR053 designed at MSU for FRIB.

| Parameters | IMP | | | MSU |
|---|---|---|---|---|
| | ES | RS | RT | RT |
| $f$ / MHz | 325 | 325 | 325 | 322 |
| $\beta_{opt}$ | 0.51 | 0.51 | 0.51 | 0.53 |
| $E_{peak}/E_{acc}$ | 4.04 | 4.78 | 3.93 | 3.53 |
| $(B_{peak}/E_{acc})$ /mT /(MV /m) | 7.07 | 7.04 | 7.49 | 8.41 |
| $(R/Q)$ /$\Omega$ | 261 | 262 | 238 | 230 |
| $G$ /$\Omega$ | 120 | 120 | 120 | 107 |

Compared with medium energy section of HWR053 that designed for FRIB at MSU, which operating at 2K, our designs have higher *G\*R/Q* that benefits to minimize the dissipated wall power for operating at 4k. Aiming to suppress the field emission effect, the design for FRIB has lowered the normalized electric field but relative higher normalized magnetic field with the race track center inner conductor. Different with 2 K system, it will produce more RF dissipated power for 4 K system, as a result of the higher BCS surface resistance than 2 K [12]. Moreover, the amount of dissipated RF power is proportional to the magnetic fields. Hence, it is more vulnerable to thermal broken, which will induce the superconducting quench, at the strong magnetic fields areas especial for 4 K system. We suppose that the lower normalized magnetic field and higher *R/Q* will likely increase the field limitation of quenching at 4 K system. However, it still needs more cryogenic test to verifying our supposition on the coming works.

## 3 Mechanical analysis and study

The cavity will be made of the high purity niobium (RRR>300) with thickness of 3.0 mm. After the deep stamping and surface preparation, the thickness of cavity is about 2.8 mm. The material property of niobium, which will be used in the simulation by ANSYS [13], are as follows: density 8570 kg /m$^3$, Poisson's ratio 0.38, Young's modulus 105 GPa, yield strength 75 MPa at room temperature and 140 MPa at 4.2 K.

The purpose of mechanical design is concentrated on keeping the static structure safety and minimizing the frequency shift caused by the helium pressure variations and Lorentz force detuning. The frequency shift is a very important behavior that must be taken into consideration for the RF generator power consumption. To maintain the electromagnetic field constant, small shift of cavity frequency will cause significant increase in the RF power of the generator due to its narrow bandwidth. For the pulse mode operation, it emphasizes on optimizing the effects of Lorentz force detuning, which dominates the behavior of the cavities. However, our prototypes are planned to operating at continue wave (CW) mode, the main perturbations contributor to the frequency shift is the fluctuation of liquid helium pressure.

### 3.1 General consideration

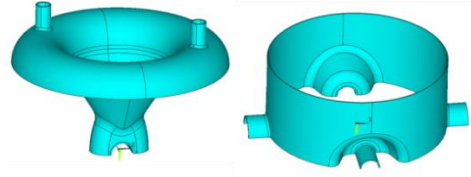

Fig. 5. The naked cavity half model can be decomposed into magnetic (left) and electric (right) domains for frequency sensitivity analysis.

The common approach to meet the requirement of structure safety and the stability of frequency, is to make the cavity more rigid by adding the stiffer ribs at the outside of wall. It is easy to decrease the equivalent stress to a certain degree if the strength of ribs is enough. However, the volume deformation cannot be avoided completely. What is more, it will be not feasible for frequency tuning if the cavity is too rigid. The sufficient method is to adjust the compensation of frequency shift contribution. Due to the Slater perturbation theory, the relationship between the frequency shifts and volume deformations are described as follows [14]:

$$\frac{\Delta f}{f} = \frac{\int_{\Delta V}(\mu|H|^2 - \epsilon|E|^2)dv}{4 \cdot U}. \quad (1)$$

Where $U$ is the total of electromagnetic energy stored in the resonator. The $\Delta f$ is the frequency shift, which the positive from high magnetic fields and the negative from high electric fields. Thus, the cavity can be divided into opposite domains, as shown in Fig. 5.

The boundary condition of the beam pipes, which connected with the tuning system, has a great distinction of frequency sensitivity. Under the two extreme conditions, for the beam pipes fixed and free, the parameters of $df/dp$ were predicted, as illustrated on the Table 2. When the pipes fixed, the total sensitivity accurately can meets with compensation from two areas. For the really operation condition, the tuner is more likely to push the beam pipes than pull them. Thus, the pipe fixed condition was chosen as our optimized target.

Table 2. The prediction of frequency sensitivity contributes from the electric and magnetic areas for the naked cavity.

| Boundary | E_areas | M_areas | $df/dp$ |
|---|---|---|---|
| Pipe fixed | 0 bar | 1 bar | +20.7 Hz/mbar |
| | 1 bar | 0 bar | -15.1 Hz/mbar |
| | 1 bar | 1 bar | +05.6 Hz/mbar |
| Pipe free | 0 bar | 1 bar | +20.7 Hz/mbar |
| | 1 bar | 0 bar | -48.2 Hz/mbar |
| | 1 bar | 1 bar | -36.9 Hz/mbar |

**3.2 Stiffener and vessel design**

To identify the location of peak equivalent stress, the pressure of 1050 mbar was load on the naked cavity. The peak equivalent stress occurred at the edge of the beam cap with the pipes fixed, as shown in the left of Fig. 6. In order to keep the tuning safety, the force of 2.5 KN applied on beam pipes. Its max equivalent stress concentrated around the welding joint between beam pipe and the cap, as shown in the right of Fig. 6. The C-shaped ribs, two daisy slabs and the helium vessel were designed, as shown in the Fig. 7.

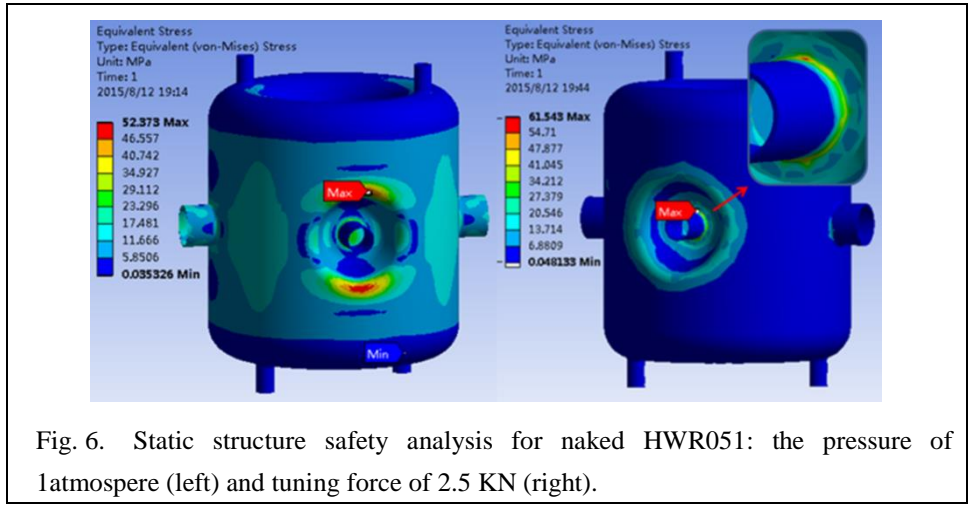

Fig. 6. Static structure safety analysis for naked HWR051: the pressure of 1atmospere (left) and tuning force of 2.5 KN (right).

The C-shaped ribs were attached at the out of beam pipes and the caps, in which the peak stress occurs, to keep the pressure safety. Generally, the more rigid of the beam caps, the more static structure safety the cavity is. However, it will be hard to move the beam pipes to obtain the wide range of frequency tuning. The C-shaped ribs can help to relief the equivalent stress and meet the requirement of tuning, as depicted in the Table 3.

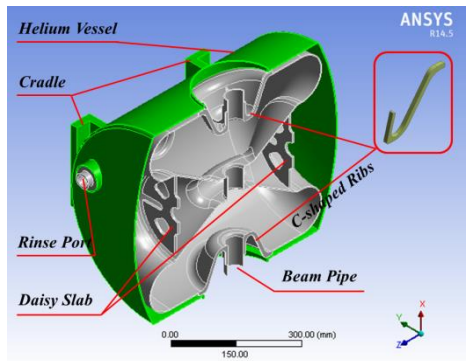

Fig. 7. Schematic diagrams of HWR051 with the ribs and helium vessel in ANSYS.

Table 3. Mechanical parameters and frequency sensitivity from higher electric areas with the C-shaped ribs.

| Peak stress/MPa | 1.0 atm | 1.5 atm | 2.0 atm | $df/dp$ Hz/mbar |
|---|---|---|---|---|
| Pipe fixed | 46.1 | 65.8 | 87.8 | -7.2 |
| Pipe free | 53.5 | 80.3 | 107 | -48.8 |

The absolute values of $df/dp$ from the electric areas will decrease from -15.1 to -7.2 Hz/mbar at the condition of pipe fixed, but it is about -48.5 Hz/mbar with no difference for the pipes free condition. The tuning sensitivity is 185 KHz/mm (total displacement) when the displacement take place on the beam pipes. The cavity stiffness is 4.9 KN/mm (total displacement) with ribs attached for tuning force on beam pipes. It is easy to get the tuning range of 160 KHz when the tuner force of 4.5 KN applied on pipes.

Above the separated analysis, the absolute value of df/dp contributes from high magnetic areas will be three times of the electric areas when the C-shaped ribs attached. The daisy slabs with thickness of 6 mm were designed on the external wall of inner conductor, where the high magnetic field around, can reduce the *df/dp* form 20.7 to 16.8 Hz/mbar with efficiency of 22%. Then, the vaulted helium vessel was optimized using titanium. Instilled on the rinsing pipes, it can furtherly decrease the df/dp to 12.8 Hz/mbar. Therefore, the total frequency sensitivity can minimalize to 5.6 Hz/mbar for the request of operation.

## 4 Conclusion

An elliptical shaped center inner conductor can get better RF properties to meet requirement of higher accelerating gradient. Meanwhile, the mechanical analysis and design reveals that the HWR051 can operate safely and stably at 4.2 K system. Two prototypes are under fabrication, and the vertical test will be conducted in the beginning of 2016.